# Period Study by the Transit Method with Ground-Based Observations

Poro A.[1,2], Elmi Kanklou F.[1,2], Ranjbaryan Iri Olya S.[2], Dashan N.[2], Ansarinia F.[2], Abdollahi F.[2], Haselpour A.[2], Dehghanizadeh Baghdadabad F.[2], Jahediparizi F.[2], Gardi A.[2], Hossein vand A.[2]

[1]The International Occultation Timing Association-Middle East section, info@iota-me.com
[2]Exoplanet Transit Project (ETP-2), Transit department, IOTA-ME, Iran

**Abstract**

In this study, 19 exoplanets were studied. Their photometric observations were obtained from the ETD. After performing the data reduction steps, its parameters were obtained through Exofast online software. Then calculated orbital parameters of planets and as a result, the periods of all the planets in this study match comparable with the Extrasolar Planets Encyclopedia. Also, we present Tc and Tdepth obtained from the observations. The location of the host stars in this study was also plotted in the H-R diagram.

**Introduction**

Throughout this study, we analyzed the orbital parameters of the following exoplanets: CoRoT-11 b, CoRoT-12 b, CoRoT-20 b, HAT-P-52 b, HAT-P-57 b, HATS-28 b, HATS-34 b, KELT-3 b, Kepler-19 b, Kepler-20 c, WASP-7 b, WASP-61 b, WASP-67 b, WASP-107 b, WASP-108 b, WASP-121 b, WASP-122 b, WASP-140 b, and WASP-163 b. All of the planets were discovered through the primary transit method, between the years 2008 and 2018. Their host stars' apparent magnitudes are between 9.51 to 15.52[1].

In order to calculate their parameters and carry out a detailed analysis, we got the raw data from the Exoplanet Transit Database (ETD[2]). Then we proceeded with data reduction use the Phoebe 0.32 software[3]. Therefore, we gave the normalized data to EXOFAST online software[4] and obtained the parameters for the planets. The purpose of this study is to make a comparison of some of the planet's parameters with other studies based on the Extrasolar Planets Encyclopedia.

**Data sets**

In this project, we got the planet's data from the Exoplanet Transit Database (ETD). ETD is one of the important and free databases of the transiting observations that is a project of the Exoplanet Section of the Czech Astronomical Society (Davoudi et al. 2020). This database is a web application that on-lined from 2008 to collect all published transit observations in one single database. This provides easy access to data for observers and exoplanet researchers. By visiting the ETD, we come across a list of exoplanets in which the number of observations of the planets and the names of their observers, as well as the observation instruments, are available. Also, the different parameters such as transit mid time, transit depth, transit duration, etc. are calculated by ETD. The ETD has three deferent parts. The first part for predictions of the transits, the next part for the observers to process and upload new data. ETD uses a simple analytical model of the transit to calculate the central time of transit, its duration, and the depth of the transit. The last part of the ETD represents plotted values in the Observed-Computed diagrams (O-C). Also, there is a parameter in the ETD called DQ (Data Quality), which grades the information of the observers according to this parameter. Their quality is graded from one to five. One symbol is the best and five symbols are the worst quality (Poddan'y et al. 2010). In Table 1 shows specifications of the host star and the planets that studied.

---

[1]http://simbad.u-strasbg.fr/simbad/
[2]http://var2.astro.cz/
[3]http://phoebe-project.org/
[4]https://exoplanetarchive.ipac.caltech.edu/cgi-bin/ExoFAST/nph-exofast

Table 1. Specifications of the host star and the planets. All of these exoplanets detected by Primary Transit method.

| Star Name | RA | Dec | Distance | Spect. | A. Mag | Planet | Discovered |
|---|---|---|---|---|---|---|---|
| CoRoT-11 | 18 42 44.94 | +05 56 15.65 | 560 | F6 V | 12.94 | CoRoT-11 b | 2010 |
| CoRoT-12 | 06 43 03.76 | -01 17 47.14 | 1150 | G2V | 15.52 | CoRoT-12 b | 2010 |
| CoRoT-20 | 06 30 52.90 | +00 13 36.85 | 1230 | G2V | 14.66 | CoRoT-20 b | 2011 |
| HATS-28 | 18 57 35.92 | -49 08 18.55 | 521 | G | - | HATS-28 b | 2016 |
| HATS-34 | 00 03 05.87 | -62 28 09.61 | 532 | - | 13.85 | HATS-34 b | 2016 |
| HAT-P-52 | 02 50 53.20 | +29 01 20.52 | 385 | - | 14.07 | HAT-P-52 b | 2015 |
| KELT-3 | 09 54 34.38 | +40 23 16.97 | 178 | F | 9.8 | KELT-3 b | 2012 |
| Kepler-19 | 19 21 40.99 | +37 51 06.43 | 219.94 | - | 12 | Kepler-19 b | 2011 |
| Kepler-20 | 19 10 47.52 | +42 20 19.29 | 290 | G8 | 12.5 | Kepler-20 c | 2011 |
| WASP-7 | 20 44 10.22 | -39 13 30.85 | 140 | F5V | 9.51 | WASP-7 b | 2008 |
| WASP-61 | 05 01 11.91 | -26 03 14.96 | 480 | F7 | 12.5 | WASP-61 b | 2011 |
| WASP-67 | 19 42 58.52 | -19 56 58.52 | 225 | K0V | 12.5 | WASP-67 b | 2011 |
| WASP-107 | 12 33 32.84 | -10 08 46.22 | 64.86 | K6 | 11.6 | WASP-107 b | 2017 |
| WASP-108 | 13 03 18.71 | -49 38 22.83 | 220 | F9 | 11.2 | WASP-108 b | 2014 |
| WASP-121 | 07 10 24.06 | -39 05 50.57 | 270 | F6V | 10.4 | WASP-121 b | 2015 |
| WASP-122 | 07 13 12.35 | -42 24 35.11 | 251.93 | G4 | 11 | WASP-122 b | 2015 |
| WASP-140 | 03 59 20.55 | -20 35 24.48 | 180 | K0 | 11.1 | WASP-140 b | 2016 |
| WASP-163 | 17 06 08.98 | -10 24 47.0 | 216.66 | G8 | 12.54 | WASP-163 b | 2018 |
| HAT-P-57 | 18 18 58.42 | +10 35 50.12 | 303 | - | 10.47 | HAT-P-57 b | 2015 |

**Observation**

Observations are gathered by different observers at various locations. All observations are made by CCDs mounted on small to medium-sized telescopes. According to Table 2, David Molina uses the smallest telescope for observing planet Kepler-20 c with an optic size of 20.3 cm and the largest optic is uses by C. Quinones and et al for observing planet WASP- 61 b with a diameter of 154 cm. The average optic size used is 46.67 cm. The most northern latitude location of observation is 64°N related to KELT-3 b planet observed by Gudmundsson and the most southern location is -32°N related to WASP-7 b planet observed by Fernández and et al. The most Eastern longitude of the location of observation is 1°E related to CoRoT-12 b planet observed by Ferran Grau Horta and the most western longitude is 296°E related to planet WASP-61 b observed by C. Quiñones and et al.

Table 2 demonstrates the data gathered by 31 observations used in this study. Including the name of the planet, name of the observer, coordinates, optic size of the telescopes, CCD type, and type of used filter. The Figure 1 shows a sample of observed and theoretical light curve.

Table 2. Observation specifications and tools

| Planet Name | Observer | Location | Optical size (cm) | CCD | Filter |
|---|---|---|---|---|---|
| CoRoT-11 b | S. Palladino, et al. | 42° N , 272° E | 60.96 | FLI ProLine 16803 | V |
| CoRoT-12 b | F. Grau Horta | 42° N , 1° E | 30.5 | CCD FLI PL1001E-1 | R |
| CoRoT-20 b | F. Lomoz | 50° N , 15° E | 30 | ST2000XM | Clear |
| CoRoT-20 b | F. Lomoz | 50° N , 15° E | 30 | ST2000XM | Clear |
| CoRoT-20 b | F. Lomoz | 50° N , 15° E | 30 | ST1200XM | Clear |
| CoRoT-20 b | F. Lomoz | 50° N , 15° E | 30 | ST2000XM | Clear |
| HAT-P-52 b | F. Campos | 41° N , 2° E | 35.56 | SBIG ST-8XME | Clear |
| HAT-P-52 b | R. Naves | 41° N , 2° E | 30.5 | CCD Moravian G4-9000 | Clear |
| HAT-P-52 b | Y. Jongen | 44° N , 5° E | 43.18 | SBIG STLX11002 | Clear |
| HAT-P-57 b | A. Wünsche | 44° N , 5° E | 82 | CCD FLI PL230 | V |
| HAT-P-57 b | F. Lomoz | 50° N , 15° E | 30 | ST2000XM | V |
| HATS-28 b | P. Evans | -21° N , 200° E | 25 | SCT | Clear |
| HATS-34 b | G. Boyle | -31° N , 149° E | 50.8 | FLI 16803 | Clear |
| HATS-34 b | Y. Jongen | -30° N , 289° E | 43.18 | Moravian 4G | Clear |
| KELT-3 b | A. Ayiomamitis | 38° N , 23° E | 30.5 | SBIG ST-10XME | Clear |
| KELT-3 b | S. Gudmundsson | 64° N , 345° E | 30 | SBIG STL 11k | Clear |
| Kepler-19 b | D. Mitchell | 35° N , 239° E | 30.8 | SBIG 10MXE CCD | Clear |
| Kepler-20 c | D. Molina | °40 N , 3° E | 20.3 | ST8xme | Clear |
| WASP-7 b | E. Fernández, et al. | -32° N , 290° E | 60 | SBIG STL 1001E | V |
| WASP-61 b | C. Quiñones, et al. | -31° N , 296° E | 154 | CCD | R |
| WASP-61 b | C. Quiñones, et al. | -31° N , 296° E | 154 | CCD | B |
| WASP-67 b | M. Bretton | 44° N , 5° E | 82 | CCD FLI PL230 | I |
| WASP-67 b | P. Evans | -21° N , 200° E | 25 | LX200 on MI-250 | Clear |
| WASP-107 b | Y. Jongen | 44° N , 5° E | 43.18 | SBIG STLX11002 | Clear |
| WASP-108 b | suricate48 | -30° N , 290° E | 35.56 | CCD | R |
| WASP-108 b | P. Evans | -21° N , 200° E | 25 | SCT | Clear |
| WASP-108 b | P. Evans | -30° N , 289° E | 35.56 | SBIG STT1603 | R |
| WASP-121 b | P. Evans | -30° N , 289° E | 35.56 | SBIG STT1603 | R |
| WASP-122 b | Y. Jongen | -30° N , 289° E | 43.18 | Moravian 4G | V |
| WASP-140 b | F. Lomoz | 50° N , 15° E | 25.4 | G2-8300 | Clear |
| WASP-163 b | A. Brosio | 39° N , 16° E | 50 | CCD | R |

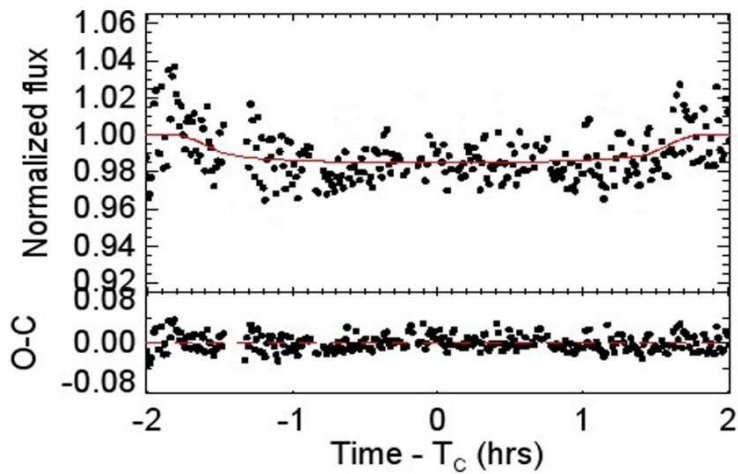

**Figure 1. Observed and theoretical light curve of HAT-P-57 b in V filter.**

**Method**

In this study, we used Exofast (Fast Exoplanetary Fitting Suite in IDL) software to obtain planetary parameters. This software is an Online Exoplanet observed data analyzer Software that works based on NASA's Exoplanet Archive (NEA). This software has been programmed in 2013 by Jason Eastman in partnership with his information team as an online analyzer robot in IDL with the California Institute of technology support (Eastman, 2013).

Collecting this archive has been started since 1995 by starting more discoveries and information of exoplanets by Mayor and Queloz the cooperating of them group and also supporting NExScI (NASA exoplanet science institute) is still in progress. This information is provided by the Kepler-the space telescope that was sent to orbit of the earth to discover Exoplanets, in 2009, by NASA, and CoRot (Convection Rotation and planetary transit- Space mission by the French national space study center) and also the given information of research observatories on the earth (Akeson, 2013).

Exofast software receives the name of the Exoplanet and observational data in the format of BJD_TDB and Flux. BJD_TDB which calculated from the online website[5], then by using Phoebe 0.32 software we calculated normalized flux.
It provides the result of analysis as an output that contains information such as the orbital period of planets, inclination, radius, and other parameters of the transit.

**Analysis and conclusion**
**A) Period accuracy**
We compared periods calculated by Exofast online software with their values on the Extrasolar Planets Encyclopedia[6]. In the Table 3, the first column is the name of the planet, the second column is the period calculated from the observations, and the third column is the period are compared. The results of observations from each planet are averaged. As can be seen, the periods of all the planets match and they are comparable.

Table 3. Compare the values of the period obtained from the observation with the Extrasolar Planets Encyclopaedia.

| Planet | Period of observation (day) | Comparison period (day) |
|---|---|---|
| CoRoT-11 b | 2.99433 | 2.994325 (± 2.1e-05) |
| CoRoT-12 b | 2.828042 | 2.828042 (± 1.3e-05) |
| CoRoT-20 b | 9.24285 | 9.24285 (± 0.0) |
| HAT-P-52 b | 2.753595 | 2.753595 (± 9.4e-06) |
| HAT-P-57 b | 2.4653 | 2.465295 (± 3.2e-06) |
| HATS-28 b | 3.181078 | 3.181078 (± 3.9e-06) |
| HATS-34 b | 2.106161 | 2.106160 (± 4.7e-06) |
| KELT-3 b | 2.70339 | 2.703390 (± 1e-05) |
| Kepler-19 b | 9.28716 | 9.286994 (± 8.8e-06) |
| Kepler-20 c | 10.854091 | 10.854091 (-2.6e-06 +3.03e-06) |
| WASP-7 b | 4.954642 | 4.954641 (± 3.5e-06) |
| WASP-61 b | 3.8559 | 3.8559 (± 3e-06) |
| WASP-67 b | 4.61442 | 4.61442 (± 1e-05) |
| WASP-107 b | 5.72149 | 5.72149 (± 2e-06) |
| WASP-108 b | 2.675546 | 2.675546 (± 2.1e-06) |
| WASP-121 b | 1.274926 | 1.274925 (± 2.5e-07) |
| WASP-122 b | 1.710057 | 1.710056 (±3.6e-06) |
| WASP-140 b | 2.235983 | 2.235983 (± 8e-06) |
| WASP-163 b | 1.609688 | 1.609688 (± 1.5e-06) |

**B) Relation between SMA and P**
We obtained the parameters from the observation of 31 exoplanet's transit, and have drawn a chart of the Semi-Major Axis of the Orbital Period. Also, we were looking for a relationship between these two parameters. According to Kepler's third formula:

$$\frac{P^2}{4\pi^2} = \frac{a^3}{G(M_1 + m_2)} \quad (1)$$

where, period (P) is given in seconds, Sami-Major Axis (a) in kilometers, and masses (M) in kilograms; these two parameters are directly related to each other, so whatever raise Semi-Major Axis, the planet's orbit is also getting bigger and the planet takes longer to orbit it's star.
Also, based on the slope of the line, this diagram 2 shows that the diagram is ascending.

---

[5] http://astroutils.astronomy.ohio-state.edu/time/
[6] http://exoplanet.eu/

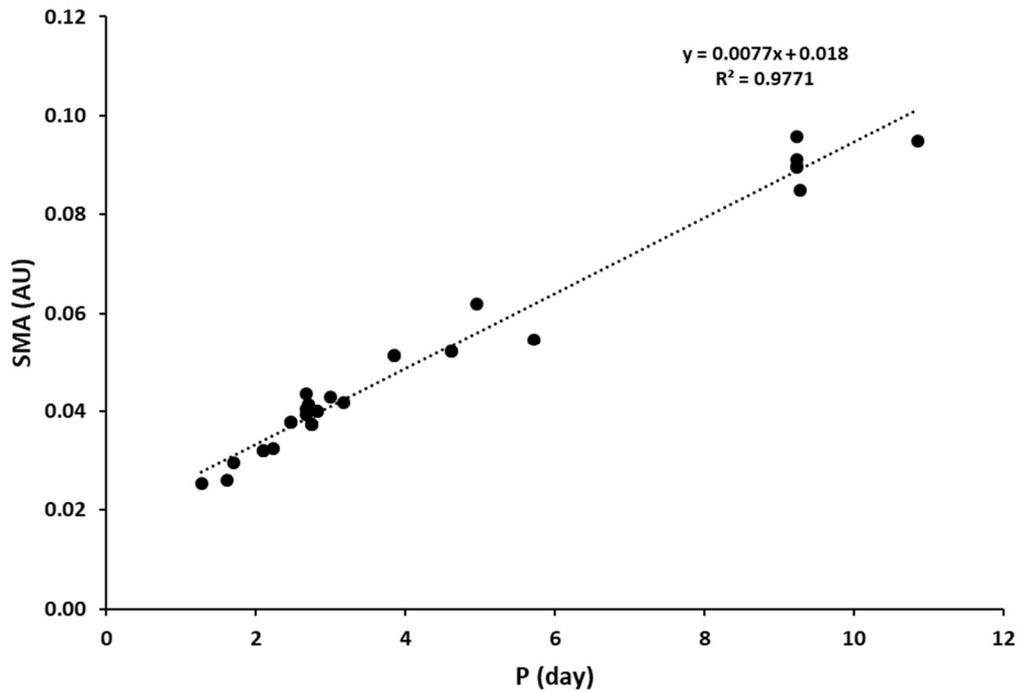

**Figure 2. Relation between SMA (AU) and P (day).**

### C) Calculating $T_c$

Time of transit, $T_c$, is the time that conjunction happens. It is represented the time of transit center and the minimum of the planet transit light curve. $T_c$ is reported in Barycentric Julian Date that is expressed as Barycentric Dynamical Time (Davoudi et al., 2019). We calculated $T_c$ for all of our observations and shows it in the Table 4.

**Table 4. Calculate the mid-time ($T_c$) of the transit for observations**

| Planet | $T_c$ | Planet | $T_c$ |
| --- | --- | --- | --- |
| CoRoT-11 b | 2456091.829404 (±0.001026) | KELT-3 b | 2456715.551831 (±0.001899) |
| CoRoT-12 b | 2457014.579814 (±0.002255) | Kepler-19 b | 2456129.893894 (±0.002722) |
| CoRoT-20 b | 2455968.465498 (±0.002736) | Kepler-20 c | 2457207.554245 (±0.002246) |
| CoRoT-20 b | 2455968.471144 (±0.002502) | WASP-7 b | 2457953.561049 (±0.001250) |
| CoRoT-20 b | 2456726.347691 (±0.002966) | WASP-61 b | 2456950.759645 (±0.001194) |
| CoRoT-20 b | 2455968.458761 ( ±0.006091) | WASP-61 b | 2456950.747158 (±0.001281) |
| HAT-P-52 b | 2458140.342804 (±0.001007) | WASP-67 b | 2458002.381915 (±0.000671) |
| HAT-P-52 b | 2457724.552899 (±0.001222) | WASP-67 b | 2456798.021280 (±0.000965) |
| HAT-P-52 b | 2458520.336274 (±0.000948) | WASP-107 b | 2458545.543719 (±0.000550) |
| HAT-P-57 b | 2458663.509091 (±0.000652) | WASP-108 b | 2458262.537298 (±0.001543) |
| HAT-P-57 b | 2458668.440058 (±0.002275) | WASP-108 b | 2457534.842474 (±0.001069) |
| HATS-28 b | 2457571.883928 (±0.001015) | WASP-108 b | 2457850.562043 (±0.000589) |
| HATS-34 b | 2457989.119324 (±0.000909) | WASP-121 b | 2457794.618947 (±0.000250) |
| HATS-34 b | 2458871.602190 (±0.001336) | WASP-122 b | 2458862.643263 (±0.000692) |
| KELT-3 b | 2457126.464382 (±0.000550) | WASP-140 b | 2457755.318812 (±0.000607) |
|  |  | WASP-163 b | 2458702.378044 (±0.001795) |

### D) Calculating T<sub>depth</sub>

We calculated the transit depth values. There are also transit depth values of the Czech Astronomical Society (ETD) are present in the Table 5. There may be several observations for each planet, an average of which has been calculated.

Table 5. Transit depth values obtained in observations and its values in ETD.

| Planet Name | Transit Depth (Observation) | Transit Depth (ETD) |
|---|---|---|
| CoRoT-11 b | 0.010352 (±0.000505) | 0.018075 (±0.00126) |
| CoRoT-12 b | 0.017022 (±0.002778) | 0.026933 (±0.00442) |
| CoRoT-20 b | 0.020805 (±0.005372) | 0.0323 (±0.0044) |
| HAT-P_52 b | 0.014011 (±0.000634) | 0.01873208 (±0.002096) |
| HAT-P_57 b | 0.010690 (±0.000539) | 0.01545 (±0.001325) |
| HATS-28 b | 0.017898 (±0.000685) | 0.0217 (±0.0011) |
| HATS-34 b | 0.018625 (±0.000935) | 0.014113333 (±0.001133) |
| KELT-3 b | 0.009705 (±0.000369) | 0.01119722 (±0.001060) |
| Kepler-19 b | 0.007244 (±0.000795) | 0.0107 (±0.0015) |
| Kepler-20 c | 0.007541 (±0.000475) | 0.0103 (±0.0013) |
| WASP-7 b | 0.016220 (±0.000188) | 0.00646666 (±0.00043) |
| WASP-61 b | 0.010062 (±0.000282) | 0.001125 (±0.00075) |
| WASP-67 b | 0.013701 (±0.000605) | 0.01645 (±0.001) |
| WASP-107 b | 0.025798 (±0.000456) | 0.03295 (±0.0007) |
| WASP-108 b | 0.015107 (±0.000431) | 0.01675 (±0.00125) |
| WASP-121 b | 0.005584 (±0.000048) | 0.0207 (±0.00055) |
| WASP-122 b | 0.014602 (±0.000320) | 0.012 (±0.000625) |
| WASP-140 b | 0.019769 (±0.000624) | 0.021575 (±0.00135) |
| WASP-163 b | 0.008744 (±0.000980) | 0.0213333 (±0.003233) |

### E) H-R diagram for the host stars

The positions of all host stars in this study, in which the theoretical Zero Age Main Sequence (ZAMS) and Terminal Age Main Sequence (TAMS), are shown in the H-R diagram in Figure 3 and the calculation are shown in Table 6. As can be seen in the diagram, most of the stars in this study are in the middle of their lifetime, and the planets were discovered at this time when star's temperature is cooler that the first part of their lifetime.

Table 6. Calculations of host stars and parameters determining their position in the H-R diagram.

| Host star | T (K) | Log T | L (L sun) | Log L |
|---|---|---|---|---|
| CoRoT-11 | 6343.0 (±72) | 3.80 | 2.31 | 0.36 |
| CoRoT-12 | 5675.0 (±80) | 3.75 | 1.30 | 0.11 |
| CoRoT-20 | 5880.0 (±90) | 3.77 | 1.58 | 0.20 |
| HAT-P-52 | 5131.0 (±50) | 3.71 | 0.66 | -0.18 |
| HAT-P-57 | 7500.0 (±25) | 3.88 | 3.85 | 0.59 |
| HATS-28 | 5498.0 (±84) | 3.74 | 0.77 | -0.11 |
| HATS-34 | 5380.0 (±73) | 3.73 | 0.85 | -0.07 |
| KELT-3 | 6304.0 (±49) | 3.80 | 2.39 | 0.38 |
| Kepler-19 | 5541.0 (±60) | 3.74 | 0.79 | -0.10 |
| Kepler-20 | 5466.0 (±93) | 3.73 | 0.72 | -0.14 |
| WASP-7 | 6400.0 (±10) | 3.81 | 2.35 | 0.37 |
| WASP-61 | 6250.0 (±15) | 3.80 | 2.00 | 0.30 |
| WASP-67 | 5200.0 (±10) | 3.72 | 0.61 | -0.21 |
| WASP-107 | 4430.0 (±12) | 3.65 | 0.27 | -0.57 |
| WASP-108 | 6000.0 (±14) | 3.78 | 1.72 | 0.24 |
| WASP-121 | 6460.0 (±14) | 3.81 | 2.88 | 0.46 |
| WASP-122 | 5730.0 (±13) | 3.76 | 3.25 | 0.51 |
| WASP-140 | 5260.0 (±10) | 3.72 | 0.69 | -0.16 |
| WASP-163 | 5499.0 (±20) | 3.74 | 0.90 | -0.05 |

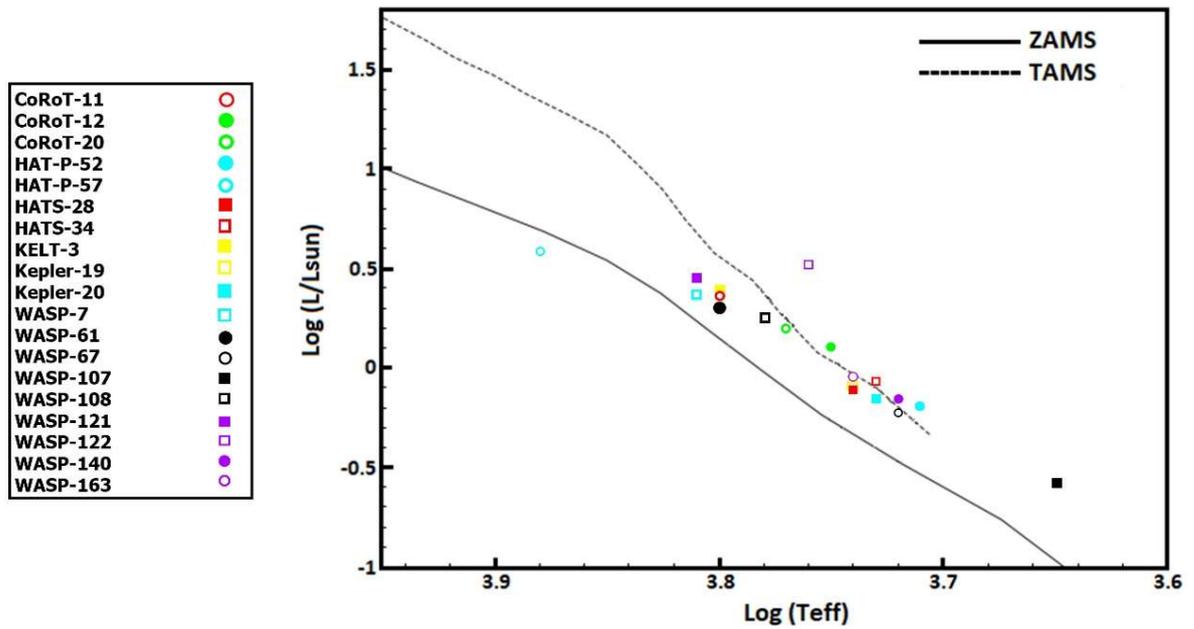

Figure 3. The position of the host stars on the H-R diagram.

## Acknowledgments
The project was held as a scientific activity on exoplanets for students by IOTA/ME in the spring of 2020.

## Appendix
Observational and theoretical light curves are included in this study, presented in the appendix.

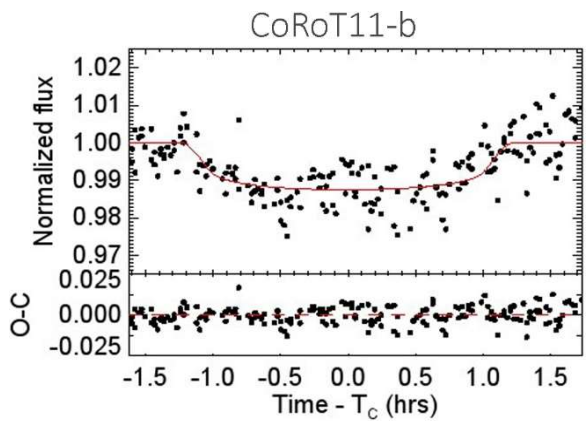
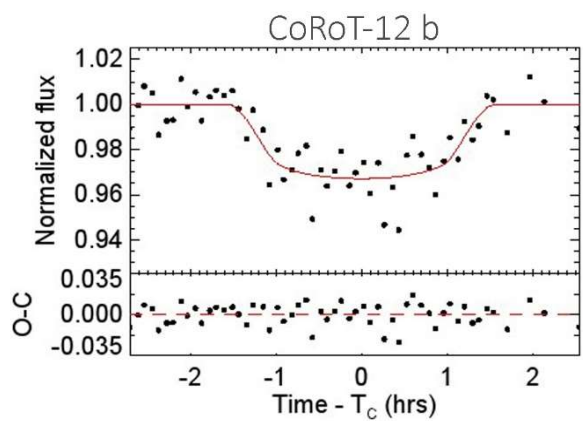
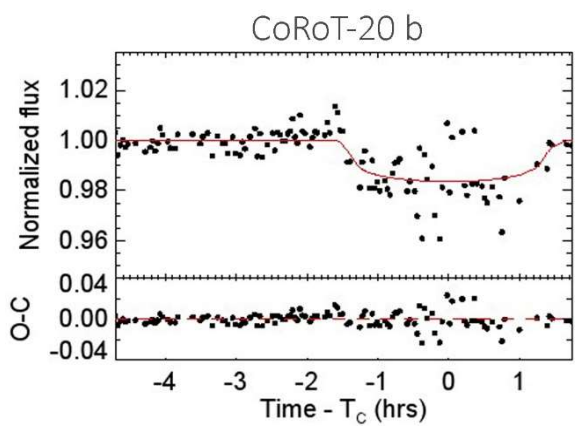
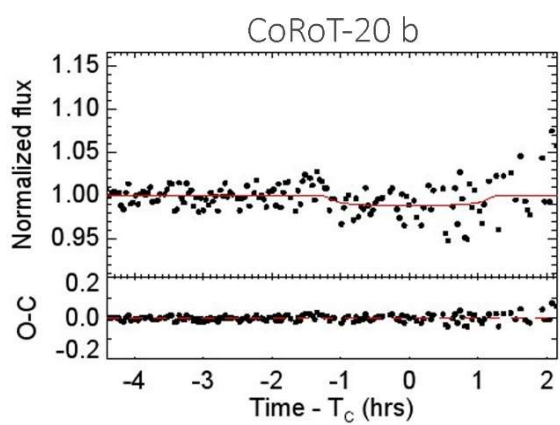
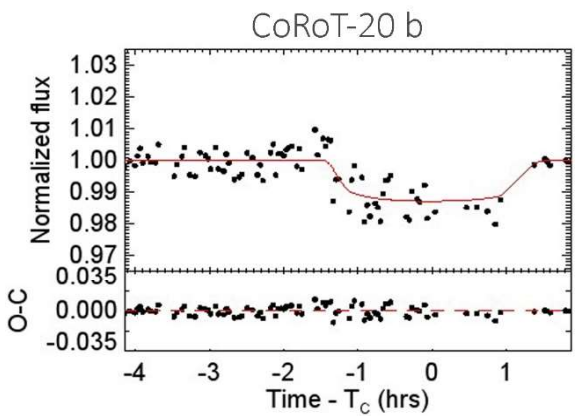
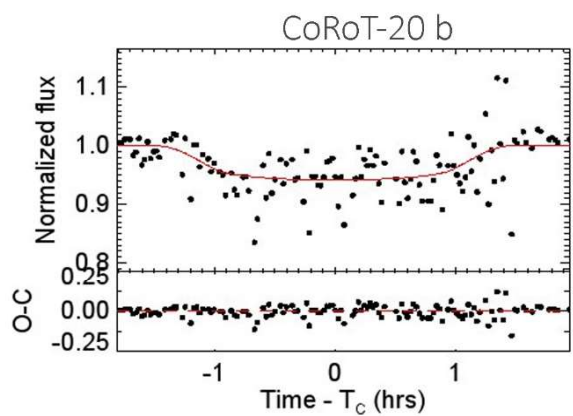
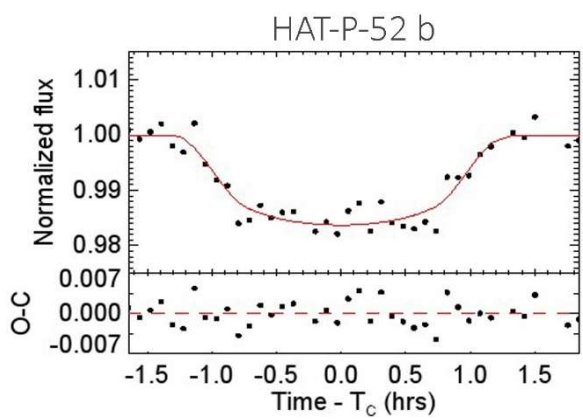
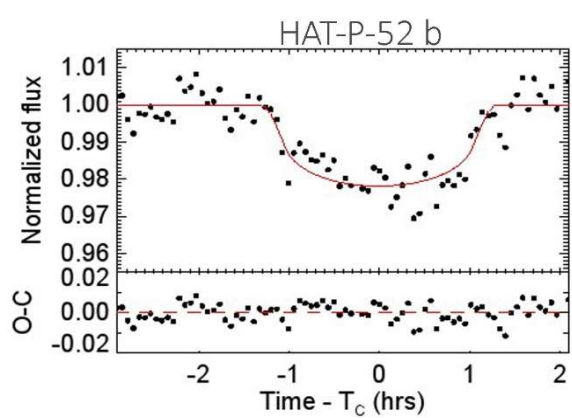

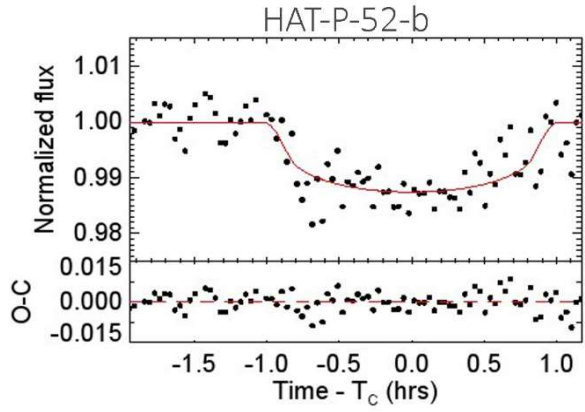
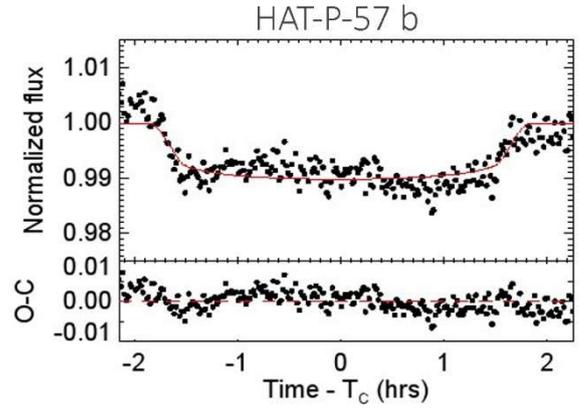
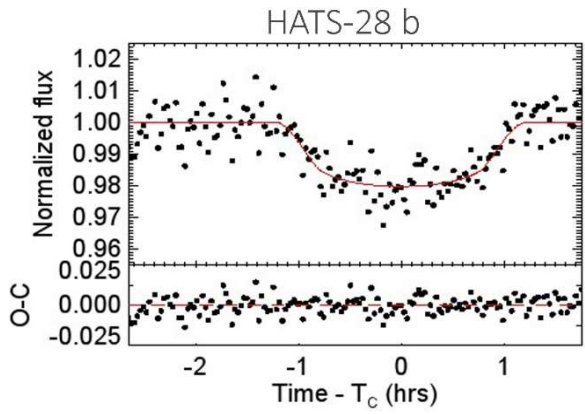
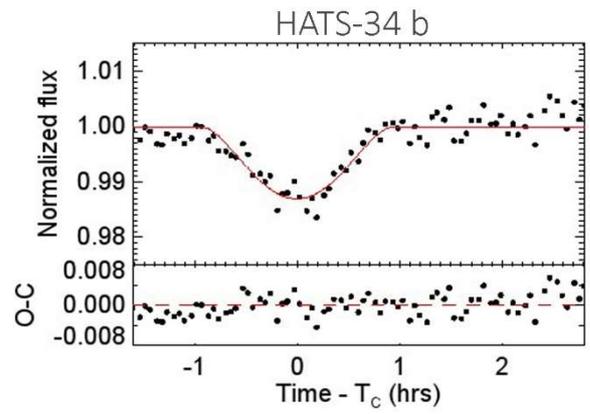
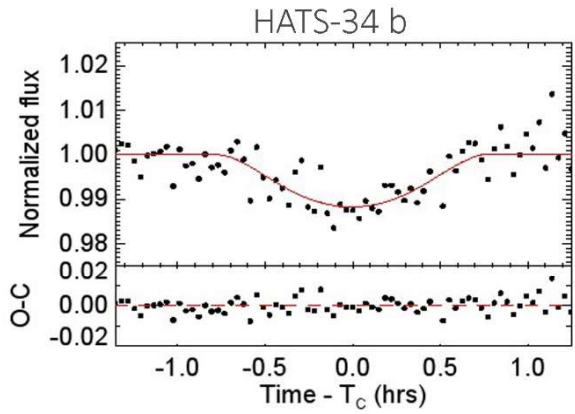
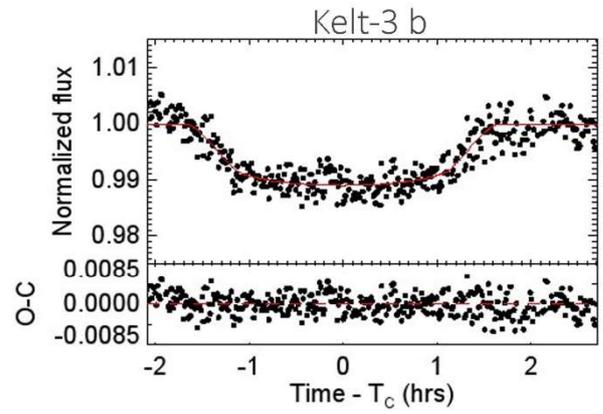
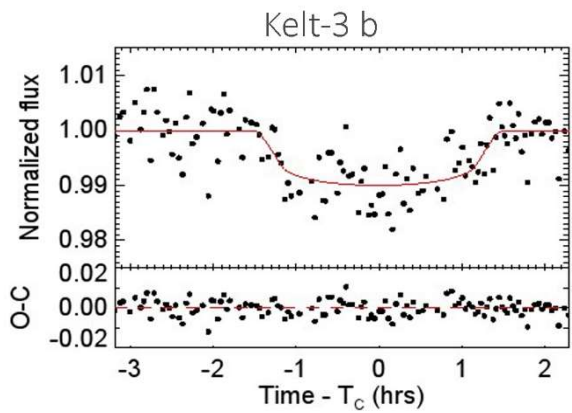
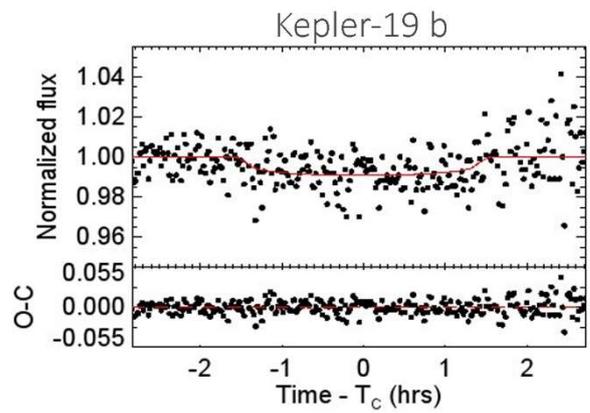

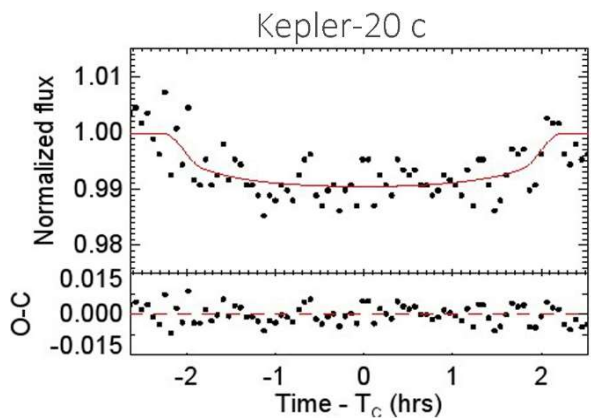
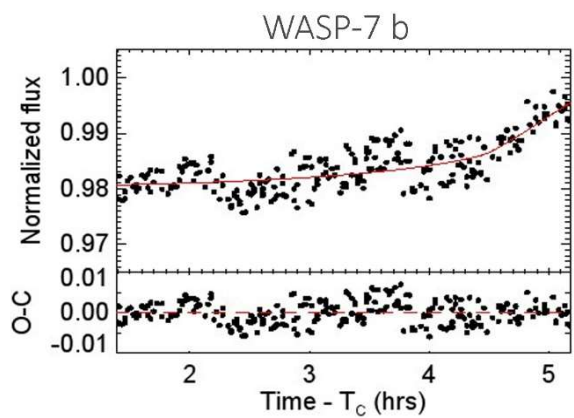
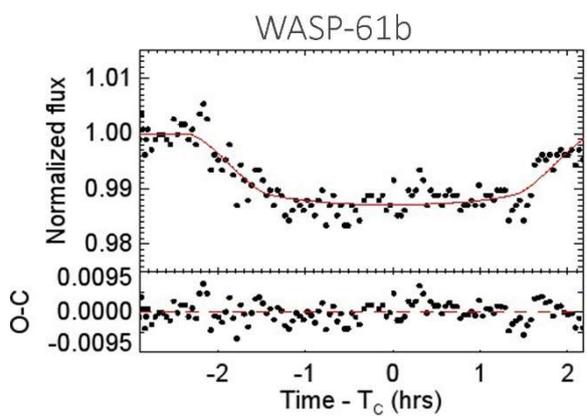
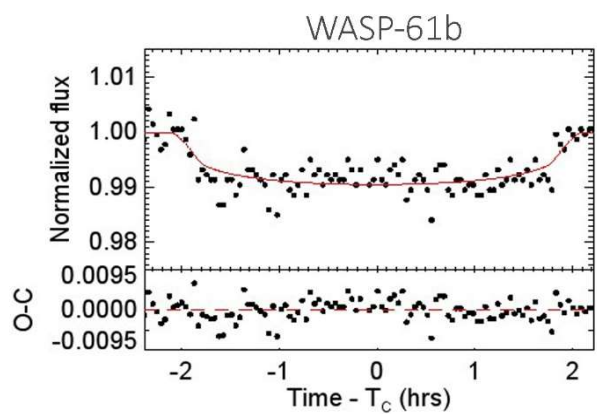
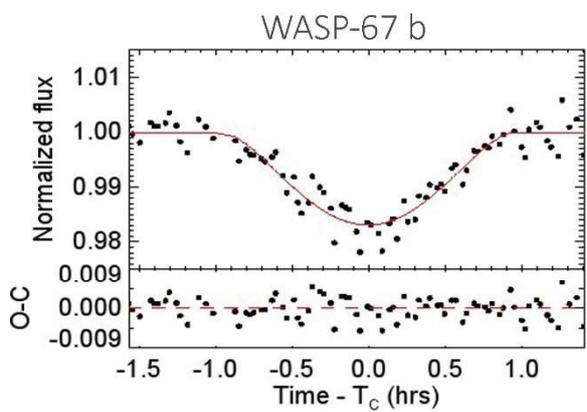
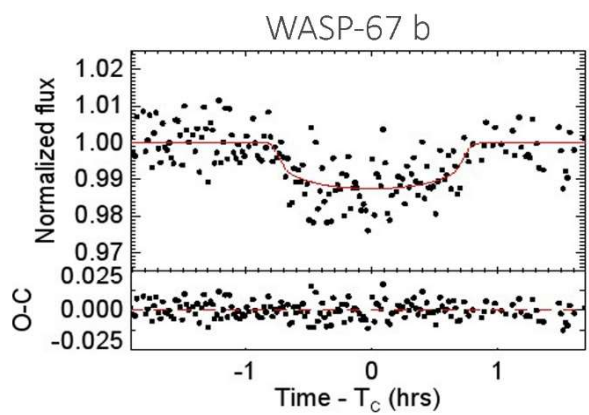
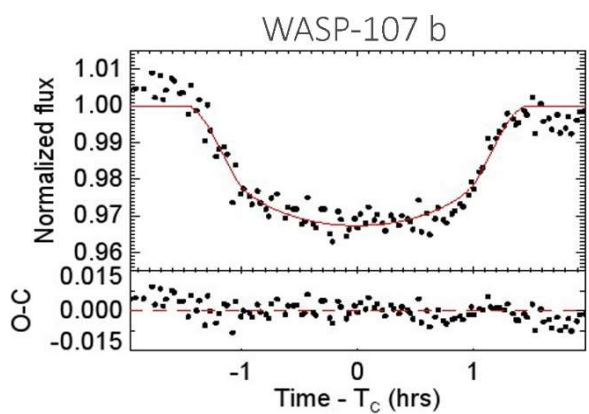
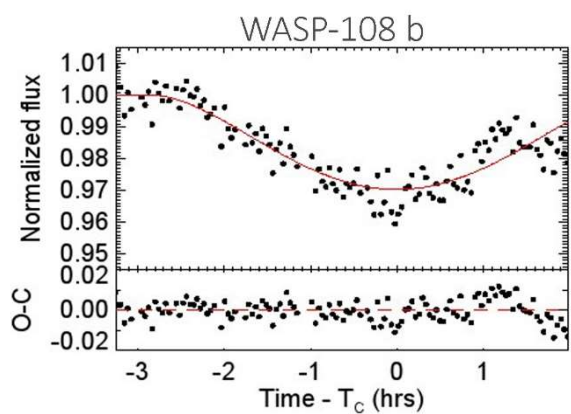

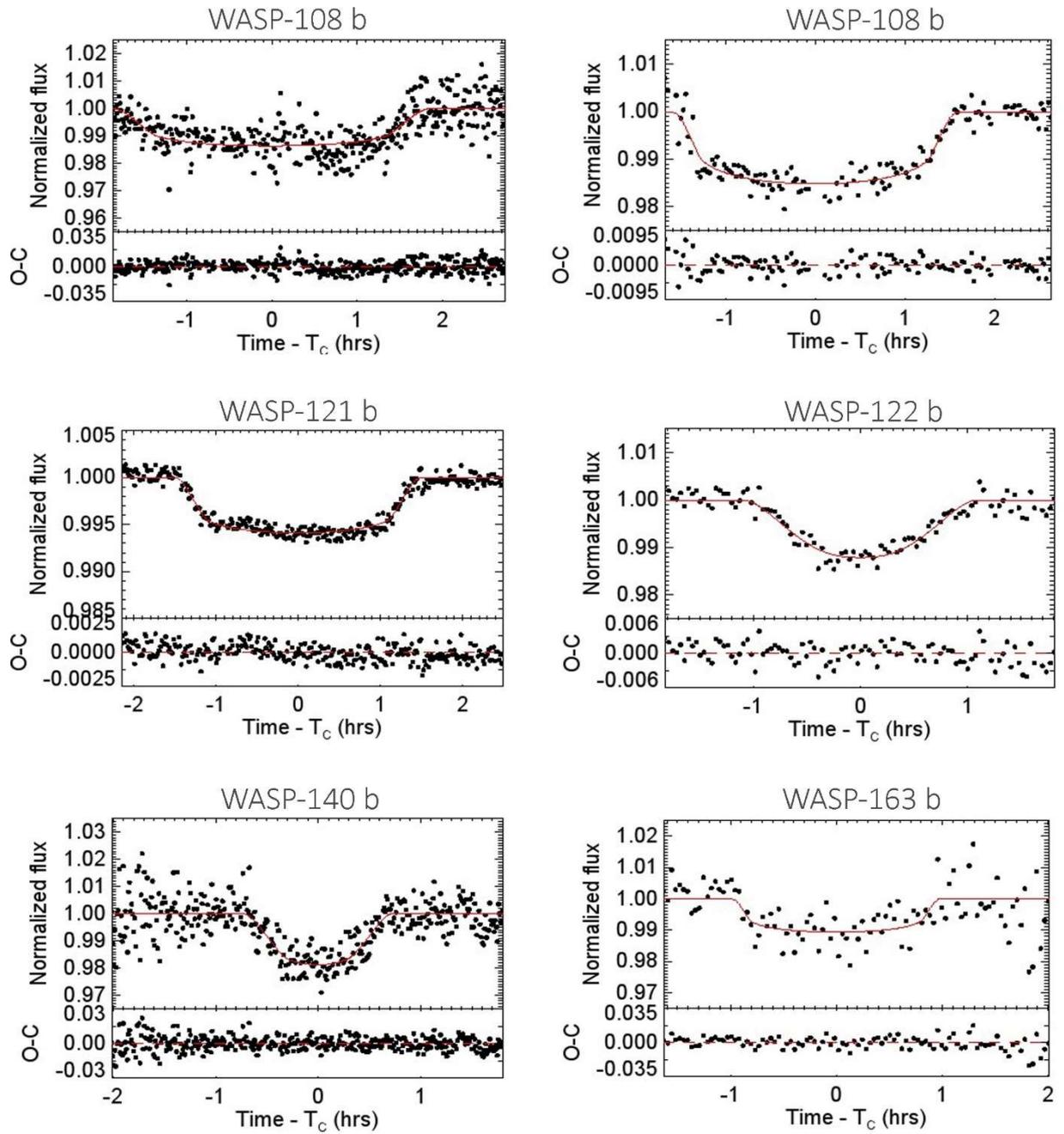